\documentclass{article}

\usepackage{PRIMEarxiv}

\usepackage[utf8]{inputenc} % allow utf-8 input
\usepackage[T1]{fontenc}    % use 8-bit T1 fonts
\usepackage{hyperref}       % hyperlinks
\usepackage{url}            % simple URL typesetting
\usepackage{booktabs}       % professional-quality tables
\usepackage{amsfonts}       % blackboard math symbols
\usepackage{nicefrac}       % compact symbols for 1/2, etc.
\usepackage{microtype}      % microtypography
\usepackage{lipsum}
\usepackage{graphicx}
\usepackage{amsmath}
\usepackage{amssymb,amsthm,amsmath}
\usepackage{xcolor,paralist,hyperref,titlesec,fancyhdr,etoolbox}
\graphicspath{{media/}}     % organize your images and other figures under media/ folder

\title{Comment on Chen et al.'s Authentication Protocol for Internet of Health Things
}

\author{
  Iman Jafarian, Siavash Khorsandi \\
  Department of Computer Engineering, Amirkabir University of Technology, Tehran, Iran \\
  \texttt{iman.j@aut.ac.ir, khorsandi@aut.ac.ir } \\
}

\begin{document}
\maketitle

\begin{abstract}
The Internet of Medical Things has revolutionized the healthcare industry, enabling the seamless integration of connected medical devices and wearable sensors to enhance patient care and optimize healthcare services. However, the rapid adoption of the Internet of Medical Things also introduces significant security challenges that must be effectively addressed to preserve patient privacy, protect sensitive medical data, and ensure the overall reliability and safety of Internet of Medical Things systems. In this context, a key agreement protocol is used to securely establish shared cryptographic keys between interconnected medical devices and the central system, ensuring confidential and authenticated communication. Recently Chen et al. proposed a lightweight authentication and key agreement protocol for the Internet of health things. In this article, we provide a descriptive analysis of their proposed scheme and prove that Chen et al.'s scheme is vulnerable to Known session-specific temporary information attacks and stolen verifier attacks.
\end{abstract}

% keywords can be removed
\keywords{cryptanalysis \and key agreement \and Internet of health things \and IoMT}

\section{Introduction}
The Internet of Things (IoT) is a revolutionary technology that has transformed the way we interact with the world around us. It encompasses a vast network of interconnected devices, sensors, and objects, all equipped with embedded sensors and actuators that enable them to collect, exchange, and act on data without requiring direct human intervention. Within the realm of IoT lies the Internet of Medical Things (IoMT), a specialized subset that integrates connected medical devices and healthcare-related applications.

The importance of security in the IoMT cannot be overstated, as it directly impacts patient safety, data privacy, and the integrity of healthcare services. Implementing a robust security protocol is a solution in the IoMT to address the unique challenges and vulnerabilities of interconnected medical devices and healthcare systems. A well-designed security protocol provides a structured and standardized framework to safeguard patient data, protect against cyber threats, and ensure the integrity and privacy of IoMT operations.

Failure to establish a secure channel in the IoMT can have serious consequences, as it may lead to unauthorized access, data breaches, and potential harm to patients. Therefore, much research has been done to provide a secure protocol in IoMT, and many articles have been published in this field in recent years [1, 2, 3, 4, 5]. In the following, we will introduce common attacks in security protocols and then review the protocol proposed by Chen et al. [6] and cryptanalyze it and show flaws in this protocol.

\section{Security Attacks}
Security protocols are designed to establish a secure and trusted communication channel but are not immune to attacks. Various attacks can be targeted against security protocols to compromise data confidentiality, integrity, and availability. In following is introduced Some common attacks on security protocols:

\begin{enumerate}
\item \textbf{Insider attack:}~Assuming that a malicious insider can access a valid user's registration information, he can easily impersonate the user by obtaining other parameters and sending data on behalf of the legal user.
\item\textbf{Stolen Mobile Device:}~If the Adversary steals the User's mobile device and recovers the datum stored in it, he can generate a valid login message to deceive the Server.
\item \textbf{Stolen Verifier Attack:}~Servers in most applications maintain verifiers of user passwords or secret keys, such as hashed passwords, instead of storing passwords or secret keys in their original clear text. With access to the verifiers, an adversary can use it to pretend to be a genuine user during a user authentication execution.
\item \textbf{Man in the Middle Attack:}~In this attack, the attacker inserts itself between a legitimate conversation and can either listen to it, modify messages or pose as one of the communication recipients.
\item \textbf{Known session key Attack:}~In this attack, It is assumed that the attacker has access to the preceding session key and then attempts to obtain the current session key.
\item \textbf{Password guessing Attack:}~The attacker tries to guess and obtain User’s password With access to transmitted parameters on the public channel.
\item \textbf{Impersonation attack:}~The adversary in an impersonation attack attempts to gain unauthorized access to information systems by masquerading as authorized users.
\item \textbf{Replay attack:}~An attack involves capturing transmitted authentication or access control information and its subsequent retransmission to produce an unauthorized effect or gain unauthorized access.
\item \textbf{Denning Sacco attack:}~The opponent is assumed to have the session key in this attack. He now tries to access long-term parameters or confidential credentials, such as user passwords or entity session keys.
\item \textbf{Known session specific Temporary information Attack:}~In this attack, the attacker tries to retrieve the session key by obtaining temporary and short-term parameters, such as the protocol's random numbers, and intercepting the parameters exchanged on the communication channel.
\item \textbf{Denial of service attack:}~In a DoS attack, the adversary forwards many illegitimate requests to block the server so legal users cannot access the server service at the right time.
\end{enumerate}

\section{Review and cryptanalysis of Chen et al.'s scheme}
\label{sec:headings}

We review and analyze the scheme of Chen et al. [6] in this section and show that this scheme suffers from session-specific known temporary information attacks and stolen verifier attacks.
\subsection{review of Chen et al.'s scheme}
The notations used in the scheme are shown in Table 1. This scheme contains two main phases: registration and authentication. In the registration phase, the user and the sensor register in the gateway. User and sensor registration phases are shown in Figure 1 and Figure 2, respectively. Then the user shares a key with the sensor via the gateway in the authentication phase—the shared key is used for their subsequent secure communications. The steps of this phase are shown in Figure 3.

\begin{table}
\scriptsize \centering \caption{Notations of Chen et al.' scheme [6]} \label{tab:PN}
\begin{tabular}{|c|c|}
\hline
\textbf{Symbol}     & \textbf{Description} \\
\hline
$U_i$        & $i$ th User\\
$ID_i$       &  Identity of $U_i$\\
$PW_i$       &  Password of $U_i$\\
$BIO$         & Biometric of $U_i$\\
$SN_j$	& $j$ th sensor node \\
$SID_j$	& Identity of $SN_j$ \\
$GWN$	& Gateway node \\
$G_j$	& Private key of $GWN$\\
$pbs$	& Public key of $SN_j$\\
$pvs$	& Private key of $SN_j$\\
$SK$		& Session key \\
$T_s$		& Timestamp, where s=1, 2, 3, 4 \\
$r_i, r_u, r_g, r_s$		& Temporary Random Numbers \\
$\oplus$		& XOR Operation\\
$||$		& Concatenation Operation\\
$hash(.)$                     & hash function\\
$Gen(.)/Rep(.)$          & fuzzy extractor/reproduction function\\
$ENC/DEC$       & Asymetric encryption/decryption\\
$ \to$               & The public channel \\
$\Rightarrow$              & The secure channel \\
\hline
\end{tabular}
\end{table}

\begin{figure}[!htbp]
\centering \scalebox{0.9}{
\begin{tabular}{|l l l|}
\hline
~~~~~~~~~~~~~~~\textbf{$U_i$}      &  ~~~~~\textbf{$Secure Channel$}    & ~~~~~~~~~~~~~~~\textbf{$GWN$}  \\
\hline
Select $ID_i, PW_i, Bio_i$               & &  \\
Select a random number $r_1$              & &  \\
Compute $HID_i=h(ID_i||r_1),$               & &  \\
~~~~~~~~~~~~~$Gen(Bio_i)=(\sigma_i,\tau_i),$               & &  \\
~~~~~~~~~~~~~$HPW_i=h(PW_i||\sigma_1),$               & &  \\
~~~~~~~~~~~~~$N=PW_i\oplus h(ID_i||\sigma_1)$               & &  \\
&         $\underrightarrow{~~~~~HID_i, HPW_i, N~~~~~}$       &  \\
& & Check $HID_i$ \\
& & Compute $D_1=h(HID_i||N)$           \\
& & Compute $D_2=h(D_1||G_j)\oplus HPW_i$\\
& & Compute $D_3=D_2\oplus N$ \\
& & Compute $D_4=h(HID_i||G_j)\oplus D_1$           \\
& & Store $\{HID_i,D_1\}$\\
& $\underleftarrow{~~~~~~~~~~D_1,D_3,D_4~~~~~~~~}$ &  \\
Compute $\Omega_i=N\oplus r_1$               & &  \\
Compute $M=h(N||r_1)\oplus HID_i$               & &  \\
Store $\{D_1,D_3,D_4,\Omega_i,M\}$               & &  \\
\hline
\end{tabular}}
\caption{User registration of Chen et al.' scheme [6]} \label{fig:UPReg}
\end{figure}

\begin{figure}[!htbp]
\centering \scalebox{0.9}{
\begin{tabular}{|l l l|}
\hline
~~~~~~~~~~~~~~~\textbf{$SN_j$}      &   ~~~~~~~~~~\textbf{$Secure Channel$}   & ~~~~~~~~~~~~~~~~~\textbf{$GWN$}  \\
\hline
~~~~~Select $SID_i$               & &  \\
&         $~~~~\underrightarrow{~~~~~~~~~~~~SID_i~~~~~~~~~~~}$       &  \\
& & Select random number $b$ \\
& & Compute $PID_j=h(SID_j||b)$           \\
& & Compute $HSID_j=h(SID_j||G_j)$\\
& & Compute $SG=h(HSID_j||G_j)\oplus PID_j$ \\
& & Compute $L=ENC_{pbs}(PID_j)$           \\
& & Store $\{SID_j,PID_j\}$\\
& $~~~~\underleftarrow{~~~~~~~~~~~~SG, L~~~~~~~~~~~}$ &  \\
~~~~~Store $\{SG, L\}$               & &  \\
\hline
\end{tabular}}
\caption{Sensor registration of Chen et al.' scheme [6]} \label{fig:UPReg}
\end{figure}

\begin{figure}[!htbp]
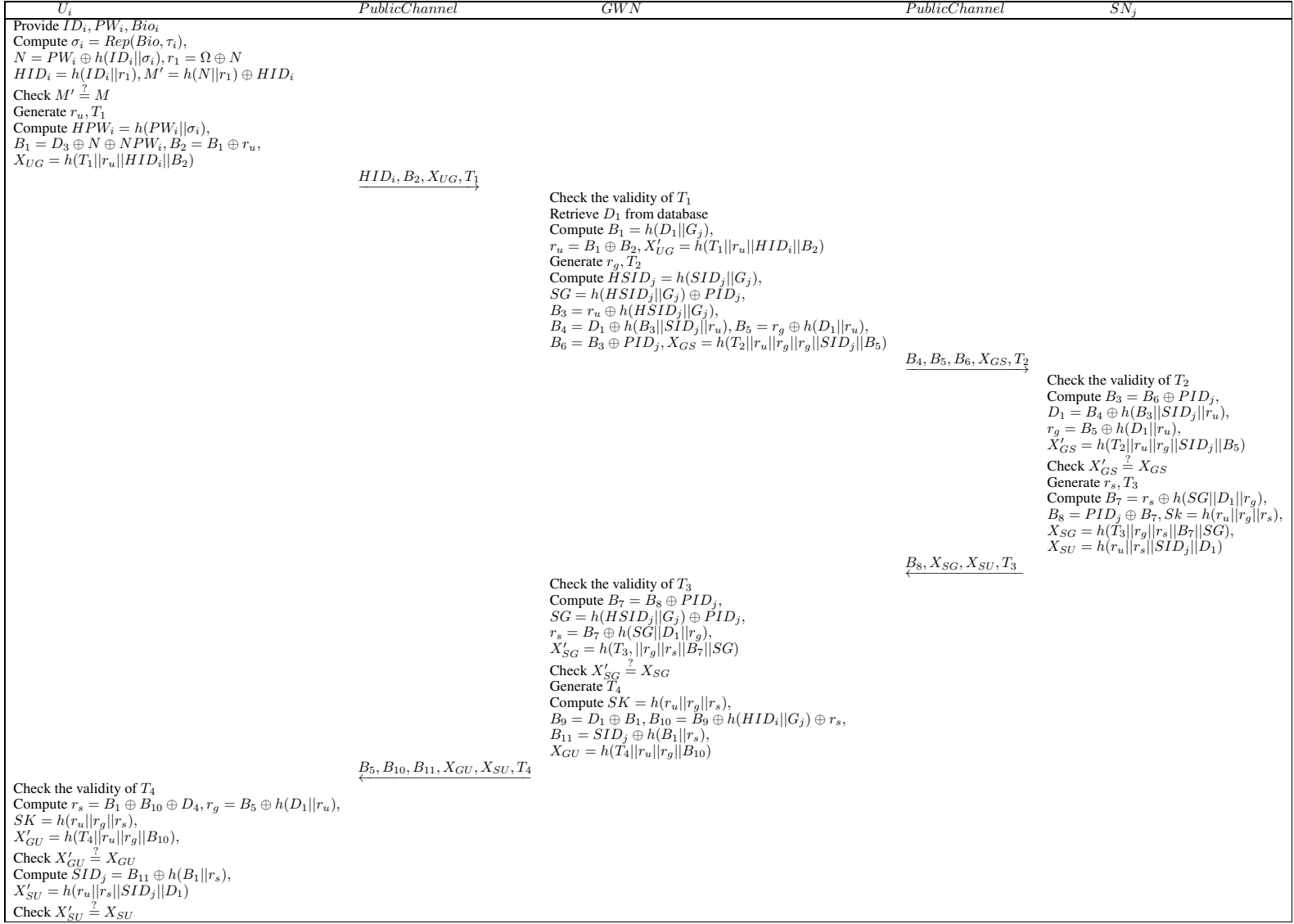

\centering \scalebox{0.68}{
\hskip-3.9cm
\begin{tabular}{|l l l l l|}
\hline
~~~~~~~~~~~~\textbf{$U_i$}&\textbf{$Public Channel$}~~~&~~~~~~~~~~~~~~\textbf{$GWN$}&\textbf{$Public Channel$}~~&~~~~~~~~~~~~~~~~ \textbf{$SN_j$}  \\
\hline
Provide $ID_i, PW_i, Bio_i$   & & & &  \\
Compute $\sigma_i=Rep(Bio,\tau_i),$   & &   & &  \\
$N= PW_i\oplus h(ID_i||\sigma_i),r_1=\Omega \oplus N$   &  &  & &  \\
$HID_i=h(ID_i||r_1),M^\prime=h(N||r_1)\oplus HID_i$   &  & & &  \\
Check $M^\prime \stackrel{?}{=} M$   & &  & &  \\
Generate $r_u, T_1$   & & & &  \\
Compute $HPW_i=h(PW_i||\sigma_i),$   & & & &  \\
$B_1=D_3\oplus N\oplus NPW_i, B_2=B_1\oplus r_u,$ & & & &  \\
$X_{UG}=h(T_1||r_u||HID_i||B_2)$   & & & &  \\
&$\underrightarrow{HID_i, B_2, X_{UG},T_1}$ & &   &   \\
& & Check the validity of $T_1$ & & \\
& & Retrieve $D_1$ from database & &          \\
& & Compute $B_1=h(D_1||G_j),$& & \\
& & $r_u=B_1\oplus B_2,X^\prime_{UG}=h(T_1||r_u||HID_i||B_2)$ & & \\
& & Generate $r_g, T_2$  & &      \\
& & Compute $HSID_j=h(SID_j||G_j),$& & \\
& & $SG=h(HSID_j||G_j)\oplus PID_j,$& & \\
& & $B_3=r_u\oplus h(HSID_j||G_j),$& & \\
& & $B_4=D_1\oplus h(B_3||SID_j||r_u), B_5=r_g\oplus h(D_1||r_u),$& & \\
& & $B_6=B_3\oplus PID_j, X_{GS}=h(T_2||r_u||r_g||r_g||SID_j||B_5)$& & \\
& &  &$\underrightarrow{B_4, B_5, B_6, X_{GS}, T_2}$   &   \\
& & & & Check the validity of $T_2$ \\
& & & & Compute $B_3=B_6\oplus PID_j,$ \\
& & & & $D_1=B_4\oplus h(B_3||SID_j||r_u),$ \\
& & & & $r_g=B_5\oplus h(D_1||r_u),$ \\
& & & & $X^\prime_{GS}=h(T_2||r_u||r_g||SID_j||B_5)$ \\
& & & & Check $X^\prime_{GS}\stackrel{?}{=}X_{GS}$ \\
& & & & Generate $r_s, T_3$ \\
& & & & Compute $B_7=r_s\oplus h(SG||D_1||r_g),$ \\
& & & & $B_8=PID_j\oplus B_7, Sk=h(r_u||r_g||r_s),$ \\
& & & & $X_{SG}=h(T_3||r_g||r_s||B_7||SG),$ \\
& & & & $X_{SU}=h(r_u||r_s||SID_j||D_1)$ \\
& &&$\underleftarrow{B_8, X_{SG}, X_{SU}, T_3~~}$ &  \\
& & Check the validity of $T_3$& & \\
& & Compute $B_7=B_8\oplus PID_j,$& & \\
& & $SG=h(HSID_j||G_j)\oplus PID_j,$& & \\
& & $r_s=B_7\oplus h(SG||D_1||r_g),$& & \\
& & $X^\prime _{SG}=h(T_3,||r_g||r_s||B_7||SG)$& & \\
& & Check $X^\prime _{SG}\stackrel{?}{=}X_{SG}$& & \\
& & Generate $T_4$ & & \\
& & Compute $SK=h(r_u||r_g||r_s),$& & \\
& & $B_9=D_1\oplus B_1, B_{10}=B_9\oplus h(HID_i||G_j)\oplus r_s,$ & & \\
& & $B_{11}=SID_j\oplus h(B_1||r_s),$& & \\
& & $X_{GU}=h(T_4||r_u||r_g||B_{10})$& & \\
&$\underleftarrow{B_5, B_{10}, B_{11}, X_{GU}, X_{SU}, T_4}$ & &   &   \\
Check the validity of  $T_4$   & & & &  \\
Compute $r_s=B_1\oplus B_{10}\oplus D_4, r_g=B_5\oplus h(D_1||r_u),$   & &   & &  \\
$SK=h(r_u||r_g||r_s),$   &  &  & &  \\
$X^\prime _{GU}=h(T_4||r_u||r_g||B_{10}),$   &  & & &  \\
Check $X^\prime _{GU}\stackrel{?}{=}X_{GU}$   & &  & &  \\
Compute $SID_j=B_{11}\oplus h(B_{1}||r_s),$   & &   & &  \\
$X^\prime_{SU}=h(r_u||r_s||SID_j||D_1)$   & &   & &  \\
Check $X^\prime_{SU}\stackrel{?}{=}X_{SU}$   & &   & &  \\
\hline
\end{tabular}}
\caption{Login and Authentication phase of Chen et al.' scheme [6]} \label{fig:UPReg}
\end{figure}

%%%%check below 
\subsection{Cryptanalysis of Chen et al.'s scheme}
In this section, we demonstrate that the scheme proposed by chen et al. [6] suffers from known session-specific temporary information attacks and stolen verifier attacks.
\subsubsection{Known session-specific temporary information attack}
Resistance against known-session-specific temporary information attack implies that if session random numbers are unexpectedly disclosed to the attacker, she should not be able to retrieve the session key $SK$. As mentioned in step 3 of the login and authentication phase, the session key $SK=h(r_u||r_g||r_s)$ depends on random numbers $r_u, r_g, r_s$ and the hash function $h$, which is public. So, if the attacker gains the random numbers $r_u, r_g$, and $r_s$, she can compute $SK$. Thus the Chen et al. scheme is vulnerable to known-session-specific temporary information attacks.

\subsubsection{Stolen verifier attack}
In the following, we demonstrate that Chen et al.’s scheme is vulnerable to stolen-verifier attacks.
During the User registration phase, user $U_i$ stores $D_1, D_3, D_4, \Omega_i$, and $M$ in his smart card memory and based on assumptions of this attack, the attacker can steal this information from the memory of the smart card.

\begin{enumerate}
\item Given that the parameter transmitted on the public channel can be intercepted and provided to the attacker. She can access the $HID_i$ that equals $M\oplus h(N||r_1)$.
\item Now the attacker can compute $HID_i\oplus M$ that obtained $M$ from the memory of the smart card, and with the help of the Self-inverse property of XOR, she can get $h(N||r_1)$.
\item Having $h(N||r_1)$ from the above step and $HID_i$ on the public channel, the attacker can create $M^\prime$, that satisfies the equation $M\stackrel{?}{=}M^\prime$.
\end{enumerate}
Hence, the attacker successfully impersonates a legitimate user for the smart card using stolen verifiers.

\section{Concludion}
Nowadays, providing a secure communication channel in the Internet of health things has been considered by many researchers. In this article, we reviewed the authentication protocol proposed by Chen et al. and demonstrated that it is prone to known session-specific temporary information attacks and stolen verifier attacks. In future, we plan to present a secure key agreement scheme for IoMT that addresses the shortcomings of related works.

%%%%%%%%%%%%%%%

\begin{thebibliography}{9}
\bibitem{1} Singh, N., \& Das, A. K. (2023). TFAS: two factor authentication scheme for blockchain enabled IoMT using PUF and fuzzy extractor. The Journal of Supercomputing, 1-50.
\bibitem{2} Xu, F., Liu, S., \& Yang, X. (2023). An efficient privacy-preserving authentication scheme with enhanced security for IoMT applications. Computer Communications.
\bibitem{3} Das, S., \& Namasudra, S. (2023). Lightweight and efficient privacy‐preserving mutual authentication scheme to secure internet of things‐based smart healthcare. Transactions on Emerging Telecommunications Technologies, e4716.
\bibitem{4} Kumar, A., Singh, K., Shariq, M., Lal, C., Conti, M., Amin, R., \& Chaudhry, S. A. (2023). An efficient and reliable ultralightweight RFID authentication scheme for healthcare systems. Computer Communications, 205, 147-157.
\bibitem{5} Servati, M. R., \& Safkhani, M. (2023). ECCbAS: An ECC based authentication scheme for healthcare IoT systems. Pervasive and Mobile Computing, 90, 101753.
\bibitem{6} Chen, C. M., Chen, Z., Kumari, S., \& Lin, M. C. (2022). LAP-IoHT: A lightweight authentication protocol for the internet of health things. Sensors, 22(14), 5401.
\end{thebibliography}
\end{document}